# Charge transfer and tunable built-in electric fields across semiconductor-crystalline oxide interfaces

Zheng Hui Lim[1], Nicholas F. Quackenbush[2], Aubrey Penn[3], Matthew Chrysler[1], Mark Bowden[4], Zihua Zhu[4], James M. Ablett[5], Tien-lin Lee[6], James M. LeBeau[3], Joseph C. Woicik[2], Peter V. Sushko[7], Scott A. Chambers[7], Joseph H. Ngai[1]

[1]Department of Physics, University of Texas-Arlington, Arlington, TX 76019, USA
[2]Materials Measurement Science Division, Material Measurement Laboratory, National Institute of Standards and Technology, Gaithersburg, MD 20899, USA
[3]Department of Materials Science & Engineering, North Carolina State University, Raleigh, NC 27695, USA
[4]Environmental Molecular Sciences Laboratory Pacific Northwest National Laboratory, Richland, WA 99352, USA
[5]Synchrotron SOLEIL, L'Orme des Merisiers, Saint-Aubin, BP 48, 91192 Gif-sur-Yvette, France
[6]Diamond Light Source, Ltd., Harwell Science and Innovation Campus, Didcot, Oxfordshire OX11 0DE, England, United Kingdom
[7]Physical Sciences Division, Physical & Computational Sciences Directorate, Pacific Northwest National Laboratory, Richland, WA 99352, USA

Built-in electric fields across heterojunctions between semiconducting materials underpin the functionality of modern device technologies. Heterojunctions between semiconductors and epitaxially grown crystalline oxides provide a rich setting in which built-in fields can be explored. Here, we present an electrical transport and hard X-ray photoelectron spectroscopy study of epitaxial $SrNb_xTi_{1-x}O_{3-\delta}$ / Si heterojunctions. A non-monotonic anomaly in the sheet resistance is observed near room temperature, which is accompanied by a crossover in sign of the Hall resistance. The crossover is consistent with the formation of a hole gas in the Si and the presence of a built-in field. Hard X-ray photoelectron spectroscopy measurements reveal pronounced asymmetric features in both the $SrNb_xTi_{1-x}O_{3-\delta}$ and Si core-level spectra that we show arise from built-in fields. The extended probe depth of hard X-ray photoelectron spectroscopy enables band bending across the $SrNb_xTi_{1-x}O_{3-\delta}$ / Si heterojunction to be spatially mapped. Band alignment at the interface and surface depletion in $SrNb_xTi_{1-x}O_{3-\delta}$ are implicated in the formation of the hole gas and built-in fields. Control of charge transfer and built-in electric fields across semiconductor-crystalline oxide interfaces opens a pathway to novel functional heterojunctions.

Charge transfer and built-in electric fields across heterojunctions between semiconducting materials underpin the functionality of device technologies that have revolutionized information processing, communication and energy harvesting. The most notable example is the pn-junction, which exhibits a built-in field between p- and n-type semiconducting materials that is fundamental to the functionality of transistors, photovoltaic devices and light-emitting diodes [1]. Similarly, charge transfer and built-in fields across heterojunctions between doped and undoped III-V semiconductors give rise to 2-dimensional electron gases that have led to discoveries such as the fractional quantum Hall effect, and devices such as high electron mobility transistors [2] [3].

Advancements in epitaxial growth now enable charge transfer and the formation of built-in electric fields to be explored across heterojunctions between crystalline oxides and semiconductors [4]. In general, semiconductor-crystalline oxide heterojunctions can exhibit functionalities beyond those of conventional semiconducting heterojunctions, given the complementary combination of covalent and ionic materials [5]. The atomically abrupt interfaces realized using oxide molecular beam epitaxy (MBE) enable the electric displacement between the oxide and semiconductor to be continuous [4] [6]. Thus far, continuity in the electric displacement between insulating dielectric oxides and semiconductors has been demonstrated, in which the surface potential of the latter can be modulated through electric fields applied through the former [4] [7] [8] [9] [10]. In some instances, electrically perfect interfaces that are virtually free of interfacial trap states have been realized [11]. In principle, such interfaces should also facilitate charge transfer between materials with itinerant carriers, which could give rise to built-in fields. The development of such heterojunctions could address emerging challenges in energy harvesting and information technology [12] [13]. Examples include heterojunctions for photocatalysis that are efficient yet also chemically stable, or extreme high-density electron gases that also exhibit high mobilities at room temperature for use in plasmonic and photonic applications. Thus, elucidating mechanisms by which charge transfer and built-in fields form across semiconductor - crystalline oxide interfaces is of fundamental and technological importance.

Here, we report charge transfer and the formation of built-in fields across heterojunctions between silicon and $SrNb_xTi_{1-x}O_{3-\delta}$ (SNTO). We find that this field is sufficiently large to induce the formation of a hole gas in the Si near room temperature for a range of Nb content *x*. The hole gas is manifested through a non-monotonic anomaly in the sheet resistance that is accompanied by a crossover in sign of the Hall resistance. Hard X-ray photoelectron spectroscopy (HAXPES) measurements reveal pronounced asymmetric features in the core-level spectra of both SNTO and Si that arise from built-in fields, as shown through modelling of the spectra. Time-of-flight secondary ion mass spectroscopy (ToF-SIMS) reveals a high concentration of oxygen impurities near the surface of the Si which enable depletion and inversion to occur. Charge transfer and built-in fields across heterojunctions between crystalline oxides and semiconductors provide another approach to electrically couple these two classes of materials and enable a new generation of functional heterojunctions.

The 12 nm thick epitaxial SNTO films were grown by oxide MBE on Czochralski-grown (001)-oriented Si substrates. The Nb content $x$ was quantified ex situ using X-ray photoemission spectroscopy (XPS) after growth. Figure 1a shows direct-space X-ray diffraction (XRD) maps of the $x = 0$, 0.084, 0.20, and 0.60, heterojunctions; the lattice parameters of SNTO and Si were determined by analyzing the (103) and (224) reflections, respectively. As the perovskite unit-cell of SNTO is rotated 45° with respect to the diamond-cubic unit-cell of Si, the lattice parameters of the former have been multiplied by $\sqrt{2}$ to enable comparison on the same plot. The SNTO films are relaxed with respect to Si for all $x$, and the lattice parameters become larger with increasing $x$. The interface between the SNTO and Si is atomically abrupt and free of secondary amorphous $SiO_x$ phases, as shown in the high-angle annular dark-field (HAADF) scanning transmission electron microscopy (STEM) image of the $x = 0.20$ heterojunction (Figure 1b).

ToF-SIMS reveals heavy oxygen impurity content (~ $10^{18}$ to $10^{20}$ cm$^{-3}$) in the near surface region of the Si wafer, as shown in Figure 1c (red). For comparison, the signal for TiO$^-$ is also shown (blue), which represents the maximum effect that knock-on could have, given the similar masses between TiO$^-$ and the Cs$^+$ ions used in the measurement. As can be seen, the concentration of O$^-$ exceeds that of TiO$^-$, in which the former is far less susceptible to knock-on given its smaller mass. Although the substrates are nominally undoped, Czochralski-grown Si inherently have oxygen impurities that can diffuse at elevated temperatures and become n-type donors [14]. As will be discussed below, we suspect these n-type donors enable depletion and the formation of a hole gas to occur.

Signatures of a hole gas forming in the Si are found in the sheet ($R_s$) and Hall ($R_{xy}$) resistances of the SNTO/Si heterojunctions. Figure 2a shows $R_s$ for the $x = 0$, 0.084, 0.20 and 0.60 SNTO/Si heterojunctions. At low temperatures, $R_s$ exhibits insulating behavior (i.e. $dR_s/dT < 0$) for $x = 0$, which progresses to metallic behavior ($dR_s/dT > 0$) as $x$ increases to 0.60. At high temperatures, non-monotonic anomalies are observed between $T \sim 265$ to $\sim 280$ K (arrows) for the $x = 0$, 0.084, and 0.20 heterojunctions, above which a sharp drop in $R_s$ is observed, followed by metallic behavior. The anomalies in $R_s$ are accompanied by non-linear behavior and a crossover in the slope of $R_{xy}$ from negative to positive with increasing temperature, as shown in Figures 2b, 2c, 2d for the $x = 0$, 0.084, 0.20 heterojunctions (symbols), respectively.

The non-linear behavior and crossover in sign of $R_{xy}$ are consistent with the emergence of parallel conduction from a hole-gas in the silicon near the interface. To quantify the sheet density and mobility of the hole-gas, we analyze the $R_{xy}$ data using a 2-carrier model that is parameterized by the sheet carrier densities $n_h$, $n_e$ and mobilities $\mu_h$, $\mu_e$ of the hole and electron carriers in the Si and SNTO, respectively (see Supporting Information) [15]. Figures 2b, 2c, 2d, and 2e show the fits (lines) to the $R_{xy}$ data for the $x = 0$, 0.084, 0.20 and 0.60 heterojunctions, respectively. The values of $n_h$ and $\mu_h$ derived from those fits are shown in Figures 2f, 2g, and 2h for the $x = 0$, 0.084 and 0.20 heterostructures, respectively. Hole sheet densities as high as $n_h \sim 3 \times 10^{12}$ cm$^{-2}$ are observed for the $x = 0.20$ heterojunction at $T > 320$ K. An average $\mu_h$ of $\sim 500$ cm$^2$V$^{-1}$s$^{-1}$ is derived from the fits for the heterojunctions that exhibit a hole gas, which is consistent with the bulk hole mobility of Si. In comparison, the corresponding values of $n_e$ and $\mu_e$ from the SNTO layers do not vary appreciably over the temperature range of 200 K $< T <$ 340 K (see Table S1). We find that only the SNTO and hole-gas in the near surface region of the Si contribute to the conductivity, as fits to the $R_{xy}$ data indicate that conductivity in the bulk of the Si substrate is negligible (Figure S1). Both the anomaly in $R_s$ and crossover in the slope of $R_{xy}$ are absent in the $x = 0.60$ sample. An upper limit to $n_h$ can be placed in the $x = 0.60$ heterojunction, as fits to the $R_{xy}$ data indicate that $n_h > 2 \times 10^{10}$ cm$^{-2}$ is not supported by the data, as shown in Figure 2e.

The emergence of a high mobility hole-gas indicates the presence of a built-in electric field across the SNTO/Si interface. We look for signatures of built-in fields using HAXPES with $\sim 6$ keV excitation, for which the probe depth exceeds the film thickness, enabling electronic information to be obtained across the buried interface [16]. Figures 3a – 3f show core-level spectra for the $x = 0$ and $x = 0.20$ SNTO/Si heterojunctions, along with reference spectra measured on bulk single crystals of SrNb$_{0.01}$Ti$_{0.99}$O$_3$ (001) and Si(001). The heterojunction Nb 3d and Ti 2p spectra show multiple features indicating formal charges ranging down to 0. We hypothesize that the lower valences are attributable to a built-in field that concentrates itinerant electrons near the SNTO/Si interface, resulting in enhanced screening of Ti$^{4+}$ and Nb$^{5+}$. We rule out inadequate oxygen flux during film growth, as the lower valences are absent in spectra for $x = 0$ and $x = 0.10$ SNTO films grown on LSAT substrates under identical oxidation and temperature conditions (Figure S2). Evidence for built-in fields is also found in the unprecedented asymmetries seen in all heterojunction core-level spectra. For photoelectron peaks originating in the film, asymmetries to high binding energy are present that are not observed in the corresponding spectra of bulk crystals with $x$ as high as 0.06, [17] or in films with $x$ up to 0.10 grown on LSAT (Figure S2). These asymmetries are indicated by arrows in Figure 3a, 3c – 3e. An asymmetry to *lower* binding energy is observed for the Si 2p spectrum (Figure 3f), which is reminiscent of an asymmetry that arises from a very thick Ti silicide layer [18]. However, STEM-HAADF imaging does not show any interfacial Ti silicide whatsoever (Figure 1b). In our analysis below, we show that the asymmetric features in the SNTO and Si core-level spectra are consistent with built-in fields across the heterojunction, and that detailed information about the spatial variations of these fields can be extracted from these data.

In order to probe the possible connection between built-in electric fields and peak asymmetries, we model Si 2p and Ti 2p spectra for the $x = 0$ SNTO/Si heterojunction (i.e. SrTiO$_{3-\delta}$/Si) using sums of spectra measured for bulk reference samples that are minimally affected by chemical shifts, surface core-level shifts and band bending (see Figure 3g). To make the Ti 2p fitting tractable, we fit the entire heterojunction spectrum and subtract all contributions due to all Ti valences except the dominant Ti 2p$_{3/2}$ feature associated with Ti$^{4+}$ and its asymmetry to higher binding energy (see Figure 3b). The fact that the asymmetry is not seen in Ti 2p spectra measured for 12 nm thick, $x = 0$ and $x = 0.10$ SNTO films grown under identical conditions on LSAT(001) (Figure S2), indicates the asymmetry is inconsistent with shake up [19]. The modelling is done by assigning the appropriate reference spectrum to each layer within the probe depth of the heterojunction. The intensities of all spectra are attenuated according to their respective depths below the surface ($z$) using an inelastic damping factor of the form exp($-z/\lambda$), in which $\lambda$ is the effective attenuation length, estimated to be $\lambda \sim 7$ nm in Si and $\sim 6$ nm in SrTiO$_3$ [16]. A built-in electric field, i.e., a gradient in the potential, will shift the binding energies of each layer relative to one another, as illustrated schematically in Figure 3h. We then fit the heterojunction spectra to sums of reference spectra over all layers by optimizing the layer-resolved binding energies.

The fitting starts by assigning randomly generated binding energies to all layers [16]. These energies are sorted and re-assigned to the layers so the binding energy at maximum intensity,

$\varepsilon_{max}(j)$, is a monotonic function of depth. This peak binding energy set $\{\varepsilon_{max}(j)\}$ is a measure of the band edge profile because core-level binding energies, like valence band maxima (VBMs), scale linearly with electrostatic potential. The spectra were then summed to generate a trial simulated heterojunction spectrum, $I_{sim}(\varepsilon)$. Optimization of the binding energies $\varepsilon$ proceeds so as to minimize a cost function, defined as

$$\chi = \sqrt{\frac{1}{n}\sum_{i=1}^{n}[I_{exp}(\varepsilon_i) - I_{sim}(\varepsilon_i)]^2 + p\sum_{j=1}^{m}[\varepsilon_{max}^k(j) - \varepsilon_{max}^k(j+1)]^2}. \quad (1)$$

The first term quantifies the goodness of the agreement between the measured and simulated spectra. The second term is designed to minimize discontinuities in the potential gradient with depth. The weighting factor $p$ is included to scale the influence of the gradient continuity condition relative to that of the spectral fit. Following the initial assignment of the binding energies to the various layers, these energies are subjected to incremental random changes and reordering. The process is repeated until $\chi$ is minimized. The superscript $k$ in Eqn. 1 indicates the order of differences between the values of the peak binding energies. The value $k = 0$ corresponds to the peak binding energies proper, whereas $k = 1$ corresponds to first differences, e.g. $\varepsilon_{max}^1(j) = \varepsilon_{max}(j) - \varepsilon_{max}(j+1)$, and so on. Here $k = 2$ is used. The two terms in Eqn. 1 are coupled. That is, increasing the parameter $p$ leads to a smoother potential profile but also to a less good fit of the simulated spectrum to experiment. We thus capped $p$ so that the first term does not exceed 0.005 for Si 2p. The same set of $k$ and $p$ parameters led to the first term being < 0.007 for Ti 2p.

The asymmetric line shapes for both Ti 2p$_{3/2}$ and Si 2p are very well reproduced by our fitting, as seen in Figure 4. For Si 2p, 350 Si layers were included in the model and the potential was varied in the first 220 of them. The contributions from deeper levels decrease exponentially and we did not observe any improvement in the quality of the Si 2p spectrum fit for $m \gtrsim 220$. All 31 layers were included and optimized for the 12 nm thick SNTO $x = 0$ film. The best-fit layer-resolved spectra are shown in Figure 4a, and the sum over layers is overlaid with the heterojunction spectra in Figure 4b. The fits are excellent in both cases.

This fitting procedure yields a spatial map of the band bending across the $x = 0$ SNTO/Si heterojunction. In Figure 4c we show the valence $E_V$ and conduction band $E_C$ edge energies as a function of distance from the interface, as extracted from the fits shown in Figure 4a-b. For both Si and SNTO $x = 0$, the valence band edge relative to the Fermi level is given by $E_V(z) = E_{CL}(z) - (E_{CL} - E_V)_{ref}$. Here $E_{CL}(z)$ is the core-level binding energy vs. $z$ and $(E_{CL} - E_V)_{ref}$ is the energy difference between the same core level binding energy and the valence band maximum measured for the pure reference material (values given below). The conduction band (CB) edge is given by $E_V(z) - E_g$, where $E_g$ is the band gap. The Si bands bend upward as the interface is approached, terminating with the VBM being degenerate with the Fermi level at the interface, thereby accommodating a hole gas, consistent with the Hall data.

The bands on the SNTO $x = 0$ side of the heterojunction also bend upward moving away from the interface, but with a somewhat smaller gradient (0.061 eV/Å) compared to the Si side (0.15 eV/Å). Continuity in the electric displacement across the interface requires that $\epsilon_{Si}E_{Si} = \epsilon_{SNTO}E_{SNTO}$, in which $\epsilon_{Si}$ ($\epsilon_{SNTO}$) and $E_{Si}$ ($E_{SNTO}$) are the dielectric constant and electric field in Si (SNTO), respectively. Taking $\epsilon_{Si} \sim 11.9$, we find that the electric displacement would be continuous for $\epsilon_{SNTO} \sim 30$, which is consistent with typical reported values of the dielectric constant of $SrTiO_3$ thin films on Si [4]. We suspect the temperature dependence of $\epsilon_{SNTO}(T)$, namely its decrease with increasing temperature [20], plays a role in the temperature dependence of the formation of the hole gas. Continuity in the electric displacement would require that a decrease in $\epsilon_{SNTO}(T)$ result in an increase in $E_{si}$, which would be consistent with the enhancement of $n_h$ that we observe.

The HAXPES, transport and ToF-SIMS measurements reveal the interplay between two key phenomena that give rise to the hole gas and built-in fields, namely, a type-III band alignment at the interface and surface depletion in the SNTO. The valence band offset (VBO) can be expressed as $\Delta E_V = \left(\Delta E_{Ti2p3-Si2p\ /2}\right)_{int} + \left(E_{Si2p3/2} - E_V\right)_{Si} - \left(E_{Ti2p3/2} - E_V\right)_{SNTO}$. Here, $\left(\Delta E_{Ti2p3/2-Si2\ /2}\right)_{int}$ is the difference between Si $2p_{3/2}$ and Ti $2p_{3/2}$ binding energies directly at the interface, and the second two terms are the differences between core-level binding energies and the VBMs for each reference material, 98.54(4) eV for Si $2p_{3/2}$ in Si(001) and 455.76(4) eV for Ti $2p_{3/2}$ in STO(001). When combined with $\left(\Delta E_{Ti2p3/2-Si2\ /2}\right)_{int} = 461.13(14) - 98.47(6) = 362.66(15)$ eV, these numbers yield a VBM = 5.44(16) eV. The CB offset ($\Delta E_C$) is given by $\Delta E_V - \Delta E_g = 3.32(16)$ eV, where $\Delta E_g$ is the difference in bulk band gaps for SNTO and Si. This type-III, or broken gap alignment, enables electrons in the valence band of Si to transfer to the SNTO conduction band, creating a hole gas in the former. We note that our analysis indicates that doping can alter band alignments at semiconductor-crystalline oxide interfaces, as prior studies of undoped SrTiO$_3$ on Si revealed a type-II band alignment [21].

Fits to the HAXPES spectra also reveal upward band bending near the surface of $x = 0$ SNTO consistent with surface depletion (Figure 4c), which is known to be pronounced in doped SrTiO$_3$ films [22]. The field induced by surface depletion propagates towards the interface and appears to be coupled to the field associated with the hole gas in the Si. If the fields associated with surface depletion and hole gas are coupled, we would expect that increasing either the thickness or carrier density of the SNTO layer would weaken the coupling, leading to a decrease in $n_h$ [22]. Indeed, transport measurements corroborate this picture, as we find that $n_h$ decreases or disappears with increasing thickness or carrier density of the SNTO layer. Figures S3a, and S3b show the $R_s$ and $R_{xy}$ data, respectively, for a $x = 0.20$ heterojunction that is 20 nm thick, i.e., 8 nm thicker than the corresponding 12 nm thick $x = 0.20$ sample considered above (Figures 2a and 2d). Fits to the $R_{xy}$ data indicate that the maximum in $n_h$ with temperature becomes nearly a factor of 10 smaller with increased thickness, as shown in Figure S3c. Similarly, the hole gas is absent in the $x = 0.60$ heterojunction (Figure 2e), which has the highest carrier density of the SNTO layers.

While the HAXPES and transport measurements implicate surface depletion and band offset in forming the built-in fields and the hole gas, other mechanisms may also contribute to the field profile in the SNTO. In particular, the density of oxygen vacancies may be enhanced near the interface for our particular growth procedure, which involved an initial anneal in ultra-high vacuum to crystalize SrTiO$_3$ (see Experimental Section).

In summary, we have demonstrated charge transfer and the formation of built-in electric fields across a heterojunction between Si and a crystalline oxide, using the carrier density in the latter as a tuning parameter. Electrical transport measurements indicate the formation of a hole gas in Si, while HAXPES enables band bending across the heterojunction to be spatially mapped. We note that techniques of band-gap engineering have recently been adapted to control band-alignments at semiconductor-crystalline oxide interfaces [8]. Control of carrier density and band alignment could enable charge transfer and built-in fields to be engineered across semiconductor-crystalline oxide heterojunctions, akin to III-V semiconductor heterojunctions. Semiconductor – oxide heterojunctions could uniquely address emerging challenges in energy harvesting and information technology, given the complementary combination of ionic and covalent materials.

**Experimental Section**

Epitaxial SNTO films were grown on (100)-oriented, nominally undoped, 2" diameter 300 μm thick Si wafers (Virginia Semiconductor) using reactive MBE in a custom-built chamber operating at a base pressure of $< 3 \times 10^{-10}$ Torr. To remove residual organics from the surface, the wafers were introduced into the MBE chamber and cleaned by exposing to activated oxygen generated by a radio frequency source (VEECO) operated at ~ 250 W. [23] Two monolayers of Sr were deposited at a substrate temperature of 550 °C, which was subsequently heated to 870 °C to remove the native layer of $SiO_x$ through the formation and desorption of SrO. A 2 × 1 reconstruction was observed in the reflection high energy electron diffraction (RHEED) pattern, indicating a reconstructed Si surface. Then a half monolayer of Sr was deposited at 660 °C to form a template for subsequent layers of STO. The substrate was then cooled to room temperature, at which 2.5 ML of SrO and 2 ML of $TiO_2$ were co-deposited at room temperature and then heated to 500 °C to form 2.5 unit-cells of crystalline $SrTiO_3$. Subsequent layers of SNTO of various thicknesses and compositions were grown at a substrate temperature of 580 °C through co-deposition of Sr, Ti, and Nb fluxes in a background oxygen pressure of $4 \times 10^{-7}$ Torr. Thermal effusion cells (Veeco and SVT Associates) were used to evaporate Sr and Ti source materials (Alfa Aesar), while Nb (Alfa Aesar) was evaporated using an e-beam evaporator (Thermionics). [23] All fluxes were calibrated using a quartz crystal microbalance (Inficon). [23] Typical growth rates were ~ 1 unit-cell per minute.

Samples for electron microscopy were prepared by conventional wedge polishing and Ar ion milling. HAADF-STEM images were acquired using a probe-corrected FEI Titan G2 60–300 kV operated at 200 kV. [23] The collection inner semi-angle and probe convergence semi-angle were approximately 77 mrad and 20 mrad, respectively. Images were acquired and processed with the revolving STEM (RevSTEM) method. [24] Each RevSTEM dataset contained 40 images of size 1024 × 1024 pixels with a 90° rotation between successive frames.

Lateral electrical transport measurements of the SNTO/Si heterojunctions as a function of temperature and applied magnetic field were performed in the van der Pauw geometry in a Quantum Design Dynacool™ System. [23] Electrical contacts were established on the four corners of diced 4 mm × 4 mm samples using Al wedge bonding (Westbond). The contacts exhibited linear characteristics in 2-point current-voltage measurements, confirming ohmic behavior. Resistivity and Hall measurements were performed using a Keithley 2400 Sourcemeter in conjunction with a Keithley 2700 fitted with a 7709 matrix module multiplexer. [23]

HAXPES measurements were made at the Diamond Light Source (UK) on the I09 Surface and Interface Structural Analysis beamline at an X-ray energy of 5930 eV using a Si(111) double crystal monochromator followed by a Si(004) channel-cut high resolution monochromator. The Scienta Omicron EW4000 high-energy hemispherical analyzer was set to 200 eV pass energy resulting in an overall experimental resolution of ~ 0.25 eV fitting the Fermi edge of a Au foil to a Fermi function. [23] The binding energy scale was calibrated using the Au 4f core levels, along with the Fermi edge of a gold foil. The x-ray angle of incidence was 5° off the surface plane, and the photoelectron detection angle was 5° off normal.

A ToF-SIMS measurement was performed at the Environmental Molecular Sciences Laboratory at Pacific Northwest National Laboratory. A TOF.SIMS5 instrument (IONTOF GmbH, Münster, Germany) was used. [23] Dual beam depth profiling strategy was used. A 1.0 keV $Cs^+$ beam (~45 nA) was used for sputtering. The $Cs^+$ beam was scanned over a 300*300 $\mu m^2$ area. A 25.0 keV $Bi_3^+$ beam (~0.57 pA) was used as the analysis beam to collect SIMS depth profiling data. The $Bi_3^+$ beam was focused to be ~5 microns diameter and scanned over a 100 × 100 $\mu m^2$ area at the center of the $Cs^+$ crater.

XPS measurements were performed at PNNL in the normal emission geometry using a Scienta Omicron R3000 analyzer and a monochromatic AlKα x-ray source ($h\nu$ = 1487 eV) with an energy resolution of ~0.5 eV. [23] The binding energy scale was calibrated using the Ag $3d_{5/2}$ core level (368.21 eV) and the Fermi level from a polycrystalline Ag foil.

**Acknowledgements**

This work was supported by the National Science Foundation (NSF) under award No. DMR-1508530. We express our gratitude to Prof. J. P. Liu for use of his Dynacool$^{TM}$ system for the cryomagnetic electrical transport measurements. XPS measurements and HAXPES fitting performed at Pacific Northwest National Laboratory were supported by the U.S. Department of Energy, Office of Science, Division of Materials Sciences and Engineering under Award #10122. The PNNL work was performed in the Environmental Molecular Sciences Laboratory, a national scientific user facility sponsored by the Department of Energy's Office of Biological and Environmental Research and located at PNNL. J.M.L. gratefully acknowledge funding from the NSF (Award No. DMR-1350273). This work was performed in part at the Analytical Instrumentation Facility (AIF) at North Carolina State University, which is supported by the State of North Carolina and the NSF (award number ECCS-1542015). The AIF is a member of the North Carolina Research Triangle Nanotechnology Network (RTNN), a site in the National Nanotechnology Coordinated Infrastructure (NNCI). We thank Diamond Light Source for access to beamline I-09 (SI17449-1) that contributed to the results presented here.

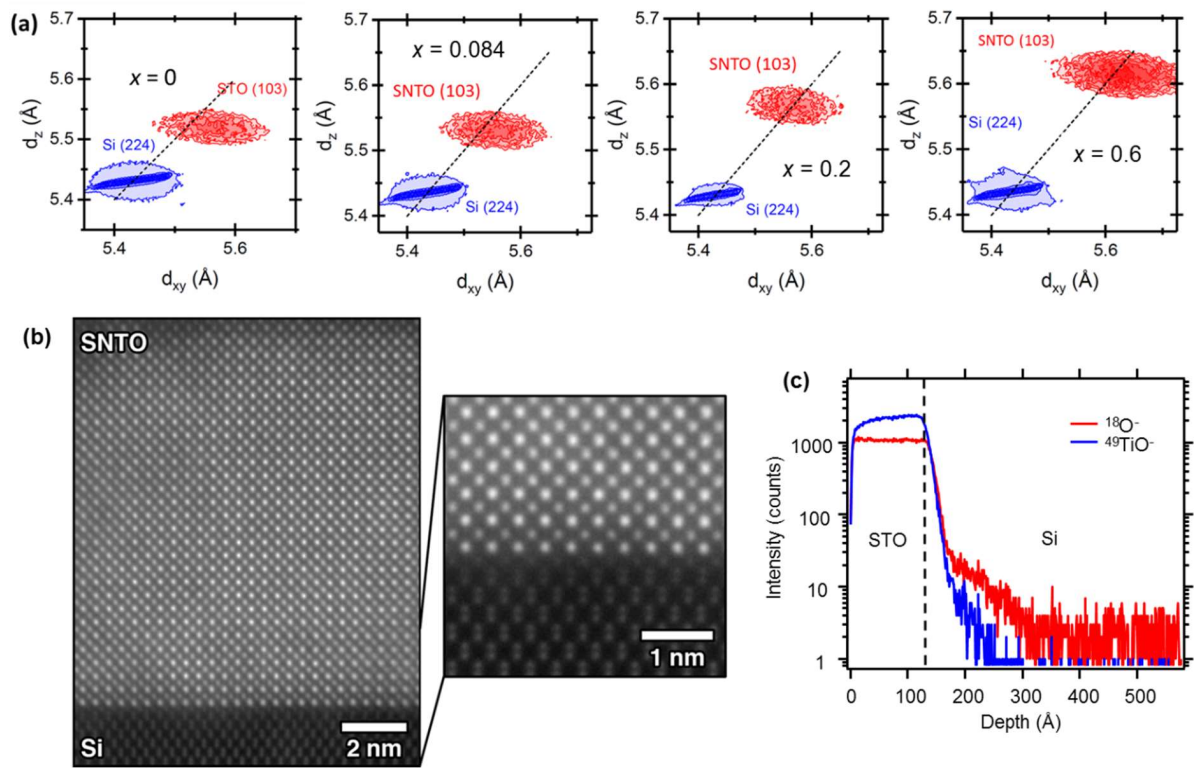

**Figure 1.** (a) Direct-space maps comparing SNTO and Si lattice parameters obtained through X-ray diffraction analysis of the (103) and (224) reflections, respectively. The SNTO films are relaxed with respect to Si, and a systematic enhancement in lattice parameters is observed with increasing $x$. (b) STEM-HAADF image of the $x = 0.20$ SNTO/Si heterojunction, showing an atomically abrupt interface. (c) ToF-SIMS showing high concentration of oxygen impurities (red) in the near surface region of Si of the $x = 0$ SNTO/Si heterojunction. Also shown is the signal for TiO$^-$ (blue), which shows the maximum effect of knock-on, given the similar masses between TiO$^-$ and the Cs ions used for the measurement.

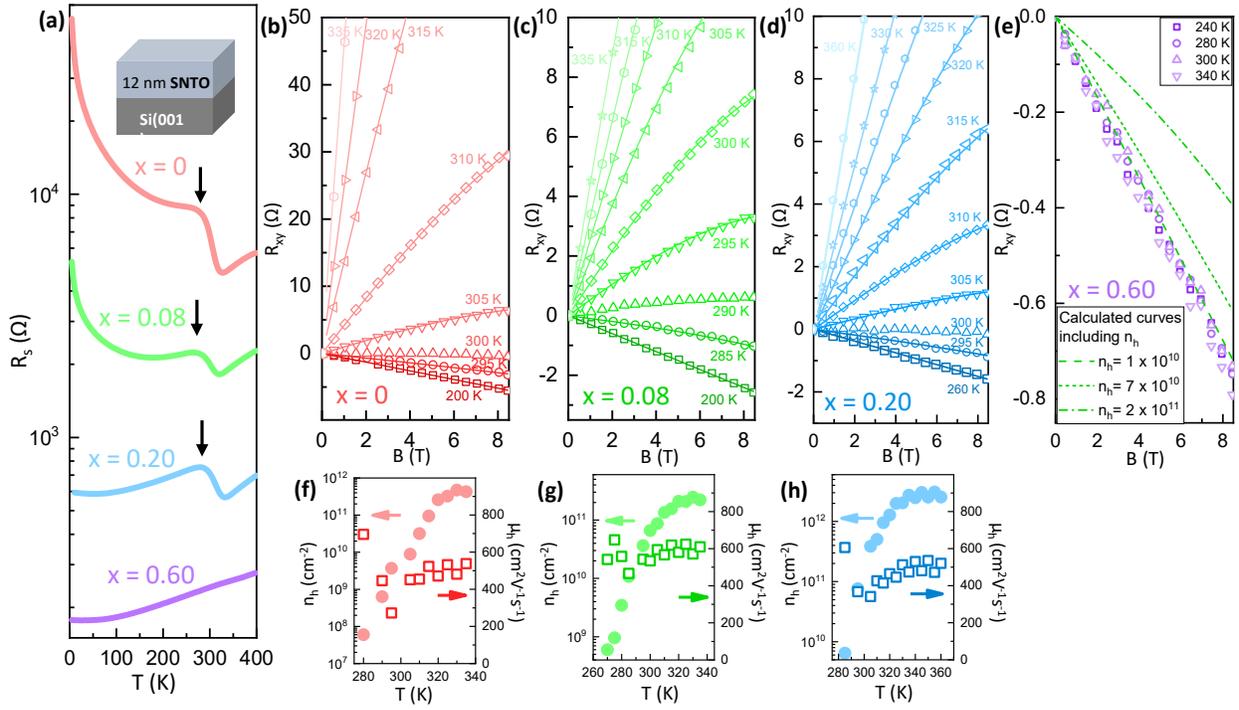

**Figure 2.** Sheet ($R_s$) and Hall ($R_{xy}$) resistances of the SNTO/Si heterostructures. (a) $R_s$ for various $x$, showing anomalies (arrows) in the $x = 0$, 0.084, and 0.20 heterojunctions. (b) to (e) $R_{xy}$ for the $x = 0$, 0.084, 0.20 and 0.60 heterostructures. Note the crossover in the sign of the slope of $R_{xy}$ for the $x = 0$, 0.084 and 0.20 heterostructures. Data are shown as symbols, while calculated fits to the data are shown as lines. (f) to (h) sheet carrier density ($n_h$) and mobility ($\mu_h$) of holes determined from fits of the $R_{xy}$ data to a 2-carrier model for the $x = 0$, 0.084 and 0.20 heterostructures.

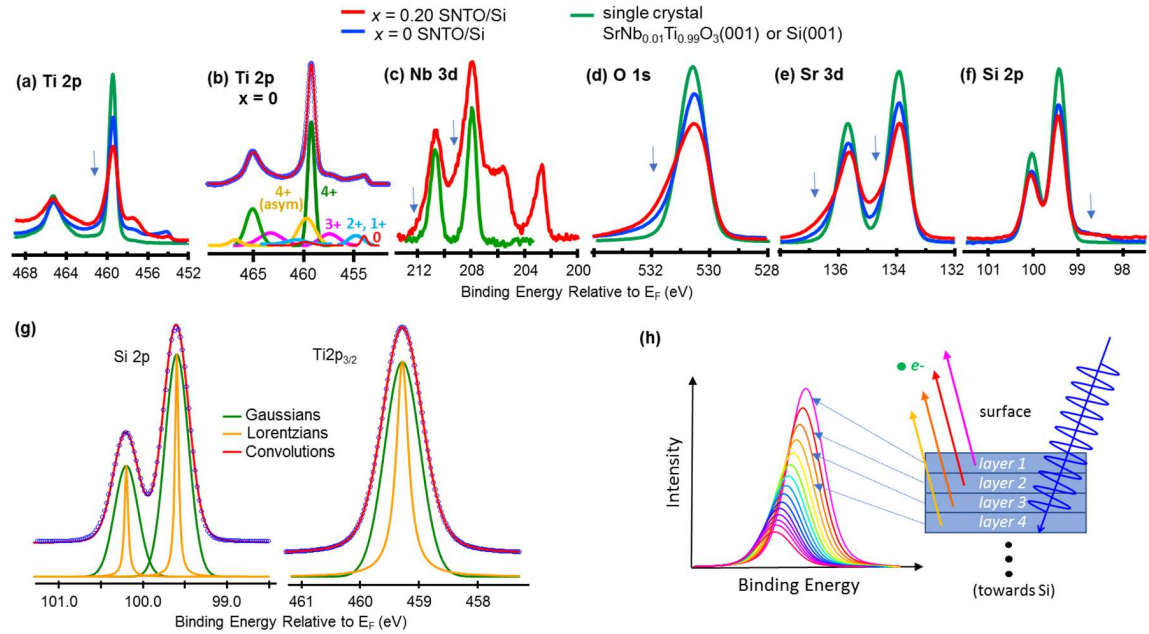

**Figure 3.** (a) – (f) Ti 2p, Nb 3d, O 1s, Sr 3d, Si 2p spectra from SNTO/Si $x$ = 0.20 (red), and $x$ = 0 (blue) heterojunctions. Spectra from a 1 at. % Nb-doped STO(001) single crystal and Si(100) substrate are also shown (green) for comparison. The Ti 2p spectra exhibit oxidation states of 0 to 4+ as shown by fits in (b). Also, note the asymmetric features observed in the core-level spectra from the heterojunctions (arrows) that are not present in the spectra of bulk substrates. (g) Reference Si 2p and Ti $2p_{3/2}$ spectra (data points) for Si(001) and 1 at. % Nb-doped STO(001) bulk single crystals, respectively. Also shown are fits to Voigt functions and the associated deconvolutions into Gaussians and Lorentzians. The excellent fits to these symmetric functions, along with the good agreement between the Lorentzian/Gaussian widths and known core-hole lifetime/experimental broadening reveals that these spectra are influenced by band bending and surface core-level shifts to a minimal extent, if at all. These spectra are thus useful for assignment to each Si or SNTO layer in the heterojunction for the purpose of extracting the band-bending induced potential profiles. (h) Schematic illustrating the method used to fit core-level spectra from the heterojunctions. A near-flat-band reference spectrum is assigned to each layer and each is attenuated in amplitude according to its depth. The binding energy of each spectrum is allowed to vary, representing the influence of a built-in electric field.

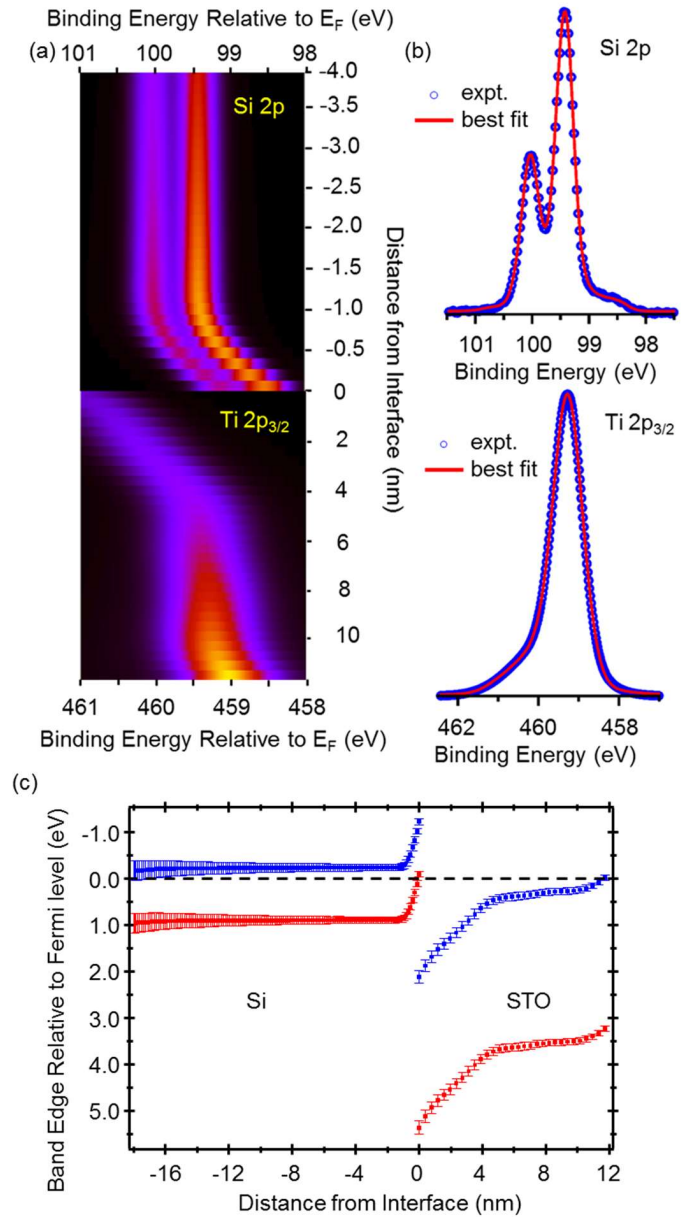

**Figure 4**. (a) Contour intensity plots of layer-resolved Si 2p and Ti $2p_{3/2}$ spectra versus distance from the interface resulting from fitting the spectra from the $x = 0$ SNTO/Si heterojunction. (b) Overlays of the sums of all spectra shown in (a) to the experimental heterojunction spectra (circles). The asymmetry to low (high) binding energy in the Si 2p (Ti$2p_{3/2}$) spectrum is interpreted as being due to the presence of large electric fields on the Si ($x = 0$ SNTO) side of the interface. (c) Band edge profiles for the $x = 0$ SNTO/Si heterojunction taken from the fits shown in (a). The conduction band edge profiles are simply the valence band edge profiles less the band gap for the appropriate material. The valence band edges were taken from the Si 2p and Ti $2p_{3/2}$ binding energy profiles corrected for the energy differences between these cores and the valence band maxima in pure reference materials.